\documentclass[twocolumn,prb]{revtex4-1}
\date{today}
\usepackage{natbib,hyperref}
\usepackage{graphicx}
\usepackage{epstopdf}
\usepackage{amsmath}
\usepackage[final]{changes}

\newcommand{\ket}[1]{\left|#1\right>}

\newcommand{\beq}{\begin{equation}}
\newcommand{\eeq}{\end{equation}}
\newcommand{\beqa}{\begin{eqnarray}}
\newcommand{\eeqa}{\end{eqnarray}}

\begin{document}
\title{Electric-dipole-induced resonance and decoherence of a dressed spin in a quantum dot}
\author{Peihao Huang$^1$}
\email{huangph@sustech.edu.cn}
\author{Xuedong Hu$^2$}
\affiliation{$^1$Shenzhen Institute for Quantum Science and Engineering, Southern University of Science and Technology, Shenzhen 518055, China\\
$^2$Department of Physics, University at Buffalo, SUNY, Buffalo, New York 14260}

\date{\today}

\begin{abstract}
Electron spin qubit in a quantum dot has been studied extensively for scalable quantum information processing over the past two decades. Recently, high-fidelity and fast single-spin control and strong spin-photon coupling have been demonstrated using a synthetic spin-orbit coupling created by a micromagnet. Such strong electrical driving suggests a strongly coupled spin-photon system. Here we study the relaxation, pure dephasing, and electrical manipulation of a dressed-spin qubit based on a driven single electron in a quantum dot. We find that the pure dephasing of a dressed qubit due to charge noise can be suppressed substantially at a sweet spot of the dressed qubit. The relaxation at low magnetic fields exhibits non-monotonic behavior due to the energy compensation from the driving field. Moreover, the longitudinal component of the synthetic spin-orbit field could provide fast electric-dipole-induced dressed-spin resonance (EDDSR). Based on the slow dephasing and fast EDDSR, we further propose a scheme for dressed-spin-based semiconductor quantum computing.

\end{abstract}

\maketitle

\section{Introduction}

Quantum information processing (QIP) employs the massive quantum parallelism from the exponentially large Hilbert space of a multi-qubit system, and is able to provide vast computing power for important tasks in physics, chemistry, engineering, etc. \cite{nielsen2010}. To take advantage of the power of QIP, and achieve fault-tolerant quantum computing or the even the near-term intermediate scale quantum (NISQ) computing, high-fidelity quantum operations on qubits are imperative \cite{fowler_surface_2012, preskill_quantum_2018}, which demands longer quantum coherence times and faster quantum gate operations \cite{nielsen2010}. 

One of the most promising candidates for QIP is an electron spin qubit in semiconductor quantum dot (QD) due to the potential scalability, fast spin manipulation, and long spin coherence time \cite{loss1998, petta2005, hanson_spins_2007, morton_embracing_2011, zwanenburg_silicon_2013, kloeffel_prospects_2013}.
Indeed, spin coherence time and the speed of quantum gates have both improved dramatically over the last decade. 
Spin coherence time has been increased from tens of nanoseconds in a GaAs QD to hundreds of microseconds in an isotope-purified silicon QD, where nuclear spin noise is reduced substantially and a ``solid-state vacuum'' is created for spin qubits \cite{petta2005, koppens_driven_2006, maune_coherent_2012, tyryshkin_electron_2003, tyryshkin_electron_2012, steger_quantum_2012, muhonen2014, veldhorst2014, veldhorst2015}. 
Spin manipulation time has been improved from microseconds to as fast as 10 nanoseconds due to the use of electric-dipole spin resonance (EDSR) through the intrinsic spin-orbit coupling (SOC) \cite{rashba_orbital_2003, flindt_spin-orbit_2006, golovach2006, nowack_coherent_2007, rashba_theory_2008, nadj-perge_spin-orbit_2010}, and especially the strong synthetic SOC (s-SOC) from a micromagnet \cite{tokura_coherent_2006, pioro-ladriere_micromagnets_2007, pioro-ladriere_electrically_2008, hu_strong_2012, yoneda_fast_2014, kawakami_electrical_2014, wu_two-axis_2014, takeda2016, yoneda2018}. Combining the long coherence time in silicon and fast spin manipulation using EDSR enabled by a micromagnet, high fidelity spin manipulation and strong spin-photon coupling have been achieved experimentally \cite{yoneda2018, zajac_resonantly_2018, watson_programmable_2018, mi_coherent_2018, samkharadze_strong_2018, xue_benchmarking_2019, borjans_resonant_2020}.

A coherently driven spin qubit can be thought of as dressed by the photons of the external electromagnetic field \cite{jing2014, laucht_breaking_2016, laucht_dressed_2017}.  Dressed states have many interesting properties.  For example, resonance fluorescence of dressed states can be minimized because of quantum interference \cite{he_dynamically_2015}. A recent experiment suggests that dephasing of an ensemble of vacancy center spins in SiC can be minimized by using dressed states, among which a clock transition exists \cite{miao_universal_2020}. Moreover, two-qubit gate can be optimized between two single- triplet (S-T$_0$) qubits by using dressed states \cite{nichol_high_fidelity_2017}. 
For an electron spin in a QD in an inhomogeneous magnetic field from a micromagnet, low frequency $1/f$ charge noise could induce strong spin relaxation and pure dephasing \cite{kha_micromagnets_2015, yoneda2018, benito_electric-field_2019, borjans_single-spin_2019, zhang_giant_2020, hollmann_large_2020, struck_low-frequency_2020, huang_impact_2020, huang_fast_2020}. It is thus intriguing to explore how dressed states of a single electron spin would behave in the presence of charge noise and inhomogeneous magnetic field, and whether dressing could mitigate decoherence and improve control.

In this work, we study relaxation, dephasing, and electric-dipole-induced resonance of a dressed-spin qubit of a single electron in a QD.
%
%
We find that phonon induced relaxation of a dressed-spin qubit is significantly modified at low magnetic fields compared to a free spin qubit due to the energy compensation from the driving field. In addition, spin dephasing due to charge noise is strongly suppressed as spin Larmor frequency approaches the driving frequency. 
Moreover, the longitudinal coupling between spin and electric field, unique to the s-SOC, leads to fast electric-dipole-induced dressed-spin resonance (EDDSR). We also propose a dressed-spin qubit based on the slow decoherence and fast EDDSR. 

\section{System Hamiltonian}
\begin{figure}[]
\includegraphics[scale=0.42]{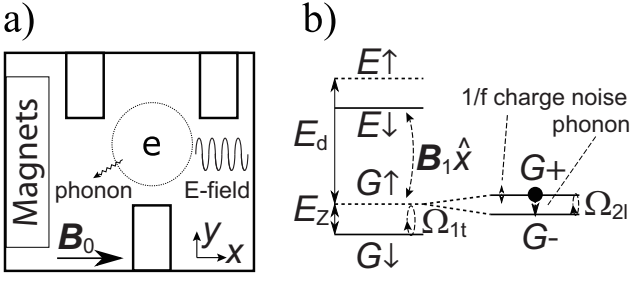}
\caption{Schematic diagrams. (a) Schematics of an electron in a gate-defined QD. Micromagnets are deposited over the QD and polarized by an external magnetic field along $x$-axis.
The nonuniform magnetic field due to the micromagnets creates a synthetic SOC that effectively couples the electron spin and the electric field, such as phonon, charge noise, or ac electric fields.
(b) Schematics of the energy levels in a QD. 
The two solid lines are orbital states, and two dashed lines represents the lifted spin degeneracies in $B_0$, where $E_d$ and $E_Z$ are the orbital and Zeeman splittings. 
The synthetic SOC $H_{SO}$ from nonuniform magnetic field hybridizes the spin-orbit states. 
When two tune ac electric field is applied, the electric fields induces $\Omega_{1t}$ and $\Omega_{2l}$ mediated by $H_{SO}$, where the transverse effective field $\Omega_{1t}$ drives the spin into dressed states, and longitudinal field $\Omega_{2l}$ provides electric-dipole-induced dressed-spin resonance (EDDSR). Electric noise, such as $1/f$ charge noise or phonon induces decoherence of the dressed  states.
}\label{schematics}
\end{figure}

We consider a single electron in a gate-defined QD [see Figure \ref{schematics} (a)], with a micromagnet deposited on top. 
The micromagnet is polarized and provides an inhomogeneous magnetic field.
Suppose the total magnetic field is $\vec{B}=\vec{B}_0+\vec{B}_1$, where $\vec{B}_0$ is a uniform magnetic field and $\vec{B}_1$ is nonuniform, 
the system Hamiltonian is thus
\begin{equation}
H= H_Z + H_d + H_{SO} +  V_{ext}(\vec{r}).
\end{equation}
Here,
$H_Z= \frac{1}{2}g \mu_B \vec{\sigma} \cdot \vec{B}_0$ arises from the Zeeman Hamiltonian of an electron spin in a uniform magnetic field $\vec{B}_0$, where $g$ is the effective g-factor, and $\vec{\sigma}$ is the Pauli operator for the electron spin.  
$H_{d} =\frac{p^2}{2m^*} + \frac{1}{2}m^* \omega_d^2 r^2$ is the 2D orbital Hamiltonian of an electron in a QD, where $\vec{r}=(x,y)$ and $\vec{p}=-i\hbar\nabla + (e/c)\vec{A}(\vec{r})$ are the 2D (in-plane) coordinate and kinetic momentum operators ($e>0$), and $\omega_d$ characterizes the in-plane confinement. The out-of-plane orbital dynamics is neglected due to the strong confinement at the interface. 
$H_{SO}$ is the s-SOC arising from the Zeeman effect in the nonuniform magnetic field $\vec{B}_1$.  For simplicity, we consider the lowest order position dependence of $\vec{B}_1$ and the inhomogeneity along x-axis, then, $\vec{B}_1=\vec{b}_1x$, where $\vec{b}_{1}\equiv \left.\partial \vec{B}_1/\partial x \right|_{x=0} \equiv [b_{1l},0,b_{1t}]$ is the gradient of the nonuniform magnetic field. Therefore, the s-SOC term is
\beqa
H_{SO} &=& \frac{1}{2}g \mu_B \vec{\sigma}\cdot \vec{b}_{1}x, \label{Hs-SOC}
\eeqa
where $\vec{b}_1$ characterizes the strength of the s-SOC.
Lastly, $V_{ext}(\vec{r})$ is the electric potential due to electrical noise or electric field for spin manipulation. As shown below, when an oscillating electric field is applied, the EDSR is achieved and drives the single spin into dressed states.

\section{Effective Hamiltonian}
We consider the limit $||H_{s-SOC}|| \ll E_Z \ll E_d=\hbar\omega_d$. To the lowest order of $H_{s-SOC}$ and $E_Z/E_d$, 
the effective Hamiltonian is \cite{huang_impact_2020} 
\beq
H_{eff}=\frac{1}{2}g\mu_B \vec{\sigma}\cdot \left(\vec{B}_0 + \vec{\Omega}\right),
\eeq
where $\vec{\Omega}=-\vec{b}_{1}\partial_x V_{ext}/(m^*\omega_d^2)$ is the effective magnetic field generated from the electric potential through QD displacement $\delta_x= -\partial_x V_{ext}/(m^*\omega_d^2)$ and the s-SOC. 
Note that $\vec{\Omega}$ can have a longitudinal component parallel to the uniform magnetic field $\vec{B}_0$, attributable to the broken time-reversal symmetry of the s-SOC \cite{huang_impact_2020}.

Consider the example when $\vec{B}_0$ is along the $x$-axis.  One can introduce a $(X,Y,Z)$ coordinate system such that the $Z$-axis is along the spin quantization axis (i.e. $x$-axis, set by the direction of $\vec{B}_0$). In the presence of an oscillating electric field $E_1(t)=E_{1max}\cos(\omega_1 t + \phi_1)$ along $x$-axis, the effective Hamiltonian becomes (let $\hbar=1$)
\beq
H_{DQ}\approx\frac{\omega_Z}{2}\sigma_Z + \vec\Omega_{1}\cdot \vec\sigma \cos(\omega_1 t +\phi_1) + \sum_{i=X,Y,Z} n_i\sigma_i, 
\eeq
where $\vec{\Omega}_1=-1/2g\mu_B\vec{b}_{1} E_{1max}/(m^*\omega_d^2)\equiv [\Omega_{1t},0,\Omega_{1l}]$ is the effective magnetic field arising from the ac electric field $E_1(t)$ mediated by the s-SOC, and $\vec{n}=-1/2g\mu_B\vec{b}_{1} E_{noise}/(m^*\omega_d^2)\equiv [n_X,0,n_Z]$ is the effective noise arising from noisy electric field $E_{noise}$. The $\Omega_{1t}$ term is responsible for the normal EDSR. In addition, the electric noise is coupled to the spin due to the s-SOC. 

In the rotating frame that rotates at the driving frequency $\omega_1$, the Hamiltonian is \cite{jing2014}
\beq
H_{DQ}^\prime=\frac{\Delta}{2}\sigma_Z + \Omega_{1t} \sigma_{X^\prime} + p_t\sigma_{X^\prime} + q_t\sigma_{Y^\prime} + n_Z\sigma_Z,
\eeq
where $\Delta=\omega_Z-\omega_1$, $p_t= \text{Re}[z_te^{-i\phi_1}]$, $q_t= \text{Im}[z_te^{-i\phi_1}]$, $z_t=(n_X+in_Y)e^{-i\omega_1 t}$. 
Note that, in the lab frame, the axes $X^\prime$ and $Y^\prime$ are rotating about $Z$-axis at the frequency $\omega_1$. However, the axes $X^\prime$ and $Y^\prime$ are static in the rotating frame.
Here we have omitted the fast oscillating terms $\Omega_{1l} \sigma_{Z^\prime}\cos(\omega_1 t)$ and $\Omega_{1t} \sigma_{+}e^{i\omega_1 t} + h.c.$ with $\sigma_\pm=\sigma_X \pm i\sigma_Y$, which can introduce coherent errors but are correctable through dynamical decoupling.

In the absence of noise, the eigenstates of $H_{DQ}^\prime$ are the dressed-spin states of interest, which is defined as the computational basis of a dressed-spin qubit.
The Hamiltonian $H_{DQ}^\prime$ without noise can be diagonalized by performing a rotation about the $Y^\prime$ axis to a new ($X^{\prime\prime}$, $Y^{\prime}$, $Z^{\prime\prime}$) coordinate system. 
After the transformation, $H_{DQ}^{\prime\prime}=U_T^\dag H_{DQ}^\prime U_T$, where $U_T= \exp(-i\theta\sigma_{Y^\prime}/2)$ with $\theta=\tan^{-1}(-2\Omega_{1t}/\Delta)$, the Hamiltonian becomes,
\beqa
H_{DQ}^{\prime\prime}&=& {\frac{\tilde\Delta}{2}\sigma_{Z^{\prime\prime}} + (p_t\cos\theta -n_Z\sin\theta)\sigma_{X^{\prime\prime}}} \nonumber\\
&& + q_t\sigma_{Y^\prime}+ (p_t\sin\theta +n_Z\cos\theta)\sigma_{Z^{\prime\prime}},
\eeqa
where $\tilde\Delta=\sqrt{\Delta^2 + 4\Omega_{1t}^2}$ is the splitting and $Z^{\prime\prime}$ is the quantization axis of the dressed-spin qubit. 
Below we show the properties of such a dressed-spin qubit, including its relaxation, pure dephasing, and ways for its manipulation.

\section{relaxation and pure dephasing of a dressed-spin qubit}
%
%
%
The quantization axis of a dressed qubit is no longer along the uniform magnetic field $\vec{B}_0$, but is determined by detuning $\Delta$ and driving amplitude $\Omega_{1t}$. Correspondingly, relaxation and dephasing of this qubit is modified compared to a free qubit \cite{jing2014}. Indeed a recent experiment suggests that a dressed qubit could provide universal coherence protection \cite{miao_universal_2020}. Here we study relaxation and dephasing of the dressed-spin qubit defined above.



To study the spin relaxation and pure dephasing, it is necessary to obtain the noise spectral density. 
For simplicity, we assume $\phi_1=0$ and noise amplitude in the $X$ and $Y$ direction are the same. Then, 
the spectral density of the noise component $p_t$ and $q_t$ is $S_{p}(\omega, \omega_1)=S_{q}(\omega, \omega_1)=\frac{1}{2}[S_{\perp}(\omega+\omega_1)+S_{\perp}(\omega-\omega_1)],$
where $S_{\perp}(\omega)$ is the spectral density of the noise along $X$ and $Y$ axis in the lab frame. 
For the noise components $p_t$ and $q_t$, the spectral density is the average of the blue- and red-shifted values from the original spectral density $S(\omega)$. 
For arbitrary detuning $\Delta$, the spectral densities in the ($X^{\prime\prime}$, $Y^{\prime}$, $Z^{\prime\prime}$) coordinate system are obtained as
$S_{X^{\prime\prime}X^{\prime\prime}}(\omega) = S_{p}(\omega,\omega_1)\cos^2\theta + S_{ZZ}(\omega)\sin^2\theta,$
$S_{Y^{\prime}Y^{\prime}}(\omega) = S_{q}(\omega,\omega_1),$
$S_{Z^{\prime\prime}Z^{\prime\prime}}(\omega) = S_{p}(\omega,\omega_1)\sin^2\theta + S_{ZZ}(\omega)\cos^2\theta.$

The relaxation $1/T_{1,DQ}$ of a dressed qubit is determined by the spectral density of noises perpendicular to the quantization axis, i.e. $1/T_{1,DQ}=S_{X^{\prime\prime}X^{\prime\prime}}(\tilde\Delta) + S_{Y^{\prime}Y^{\prime}}(\tilde\Delta)$. Therefore,
\beq
1/T_{1,DQ}= S_{p}(\tilde\Delta,\omega_1)(1+\cos^2\theta) + S_{ZZ}(\tilde\Delta)\sin^2\theta, \label{T1DQ}
\eeq
where $S_{p}(\tilde\Delta,\omega_1)$ is the modulated spectral density obtained above. 
The spectral density $S_{ZZ}(\tilde\Delta)$ depends on the noise along $Z$-axis, which is proportional to longitudinal field gradient $b_{1l}$. 

There is also pure dephasing of a dressed qubit due to phonon noise. The pure dephasing $1/T_{\varphi,DQ}$ of the dressed qubit depends on the spectral density along the $Z^{\prime\prime}$-axis. For a dressed qubit, the effective spectral density $S_{Z^{\prime\prime}Z^{\prime\prime}}(\omega)$ of phonon noise becomes finite at zero frequency due to the energy compensation from the driving field. 
Suppose phonon noise is Markovian, then, the dephasing is given by $S_{Z^{\prime\prime}Z^{\prime\prime}}(\omega=0)$.
Thus, the pure dephasing $1/T_{\varphi,DQ,ph}$ of a dressed qubit due to phonon is 
\beq
1/T_{\varphi,DQ,ph}=\frac{1}{2}S_{\perp}(\omega_1)\sin^2\theta. \label{Tphiph}
\eeq

The dephasing from the $1/f$ charge noise has been studied previously for a static qubit without driving \cite{huang_impact_2020}. For a dressed-spin qubit, the pure dephasing due to the quasi static $1/f$ charge noise is $1/T_{\varphi,DQ,1/f}\propto \sqrt{S_{Z^{\prime\prime}Z^{\prime\prime}}(\omega=1)}$ \cite{hu2006,huang_spinDephasing_2018}. Note it is negligible for the contribution of $1/f$ charge noise to $S_{p}(\omega,\omega_1)\sin^2\theta$ term in $S_{Z^{\prime\prime}Z^{\prime\prime}}(\omega)$ due to the shift of the center frequency, then, the pure dephasing of a dressed-spin qubit from $1/f$ charge noise is 
\beq
1/T_{\varphi,DQ,1/f}=1/T_{\varphi,1/f}|\cos\theta|, \label{Tphi1ovf}
\eeq
where $1/T_{\varphi,1/f}$ is the dephasing due to $1/f$ charge noise of a non-driven spin qubit studied previously \cite{huang_impact_2020}.
$1/T_{\varphi,DQ,1/f}$ is proportional to that for a static qubit but scaled by a factor $|\cos\theta| \sim |\Delta|/\tilde\Delta$, so that $1/T_{\varphi,DQ, 1/f}$ is slower than the case of a static spin qubit. 
Below, we report the numerical values of the relaxation and pure dephasing of a dressed qubit due to $1/f$ phonon emission and $1/f$ charge noise.

\begin{figure}[t]
\includegraphics[scale=0.35]{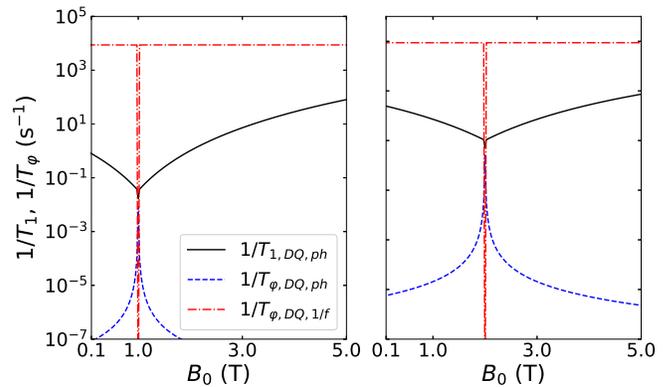}
\caption{Pure dephasing $1/T_{\varphi,DQ}$ and relaxation $1/T_{1,DQ}$ of a dressed-spin qubit as a function of the magnetic field $B_0$ due to $1/f$ charge noise and phonon when driving field frequency $\omega_1=\omega_Z(B_0=1)$ (left) or $\omega_1=\omega_Z(B_0=2)$ (right), 
where $\omega_Z/(2\pi)$ is the Larmor frequency in a magnetic field $B_0$.
Black solid, blue dashed, and red dash-dotted lines are relaxation $1/T_{1,ph}$ due to phonon, and pure dephasing $1/T_{\varphi,ph}$ and $1/T_{\varphi,1/f}$ due to phonon and $1/f$ charge noise.
Relaxation $1/T_{1,ph}$ of a dressed qubit reaches minimum when $\omega_Z(B)=\omega_1$, and scales as $B_0^5$ when $B_0>\omega_1/(g\mu_B)$. Pure dephasing $1/T_{\varphi,ph}$ due to phonon maximizes when $\omega_Z =\omega_1$.
While pure dephasing $1/T_{\varphi,1/f}$ due to $1/f$ charge noise is dramatically reduced for a dressed qubit when $\omega_Z =\omega_1$.}\label{T1invDQ}
\end{figure}

Figure \ref{T1invDQ} shows the spin relaxation $1/T_{1,DQ}$ of a dressed qubit as a function of the magnetic field $B_0$ when the microwave frequency $\omega_1=\omega_Z(B_0=1)$ (left panel) and $\omega_1=\omega_Z(B_0=2)$ (right panel).
The spin relaxation $1/T_{1,DQ}$ shows a transition when $g\mu_B B_0 = \hbar\omega_1$. The spin relaxation first decreases and reaches a minimum value at the transition point, then, it increases as $B_0^5$. The relaxation $1/T_{1,DQ}$ when $B<\hbar\omega_1/(g\mu_B)$ is bigger than the relaxation rate at the transition point, which is due to the fact that when $B<\hbar\omega_1/(g\mu_B)$, the bare splitting of the dressed qubit is approximately $\Delta=|\omega_Z-\omega_1|$, thus, the spin relaxation is approximately the spin relaxation when $\omega_Z^\prime-\omega_1=|\Delta|=\omega_1-\omega_Z$, or $\omega_Z^\prime=2\omega_1 - \omega_Z$. Furthermore, there are sharp dips at the transition points, which are due to the suppression of the coefficient $\cos^2\theta$ in the first term in Eq. (\ref{T1DQ}). The spin relaxation is non-zero at the transition point, where $\cos\theta=0$, because of the finite contribution of $S_{Y^{\prime}Y^{\prime}}(\tilde\Delta)=S_p(\tilde\Delta,\omega_1)$. 

Figure \ref{T1invDQ} also shows the spin pure dephasing $1/T_{\varphi,DQ}$ of a dressed qubit due to phonon emission and $1/f$ charge noise. 
The pure dephasing $1/T_{\varphi,DQ,ph}$ due to phonon emission maximizes when $g\mu_B B_0=\hbar\omega_1$, where $1/T_{\varphi,DQ,ph}$ is less than the 10 $s^{-1}$.
Away from the maximum point, $1/T_{\varphi,DQ,ph}$ is reduced due to the reduced hybridization, where the coefficient $\sin^2\theta$ in Eq. (\ref{Tphiph}) is reduced. 
The maximum dephasing rate $1/T_{\varphi,DQ,ph}$ ($B_0=\hbar\omega_1/g\mu_B$) reduces when the driving frequency $\omega_1$ is reduced (see panels in Figure \ref{T1invDQ}), since the dephasing is proportional to $S_{\perp}(\omega_1)$, which scales as $\omega_1^5$. 

Figure \ref{T1invDQ} further shows the pure dephasing $1/T_{\varphi,DQ,1/f}$ due to $1/f$ charge noise.  $1/T_{\varphi,DQ,1/f}$ is almost the same as the case of a non-driven spin qubit, when the spin Zeeman splitting is off resonance with the driving field $\omega_1 \ne \omega_Z$. However, at the sweet spot when $\omega_1 =\omega_Z$, the dephasing from the $1/f$ charge noise to the lowest order is strongly suppressed.
Thus, the coherence time may have more than hundreds of times increase when $\omega_1 = \omega_Z$ (the higher order dephasing process is negligible when the driving is weak), which makes a dressed-spin qubit favorable for quantum computing applications.

\section{EDDSR and "ultra-strong" driving of a dressed qubit}
%
In a conventional spin resonance experiment, a constant magnetic field establishes the quantization axis of the spins, while a small transverse AC magnetic field is used to flip the spins. The same principle is behind EDSR experiments (and our considerations so far in this manuscript) in the context of s-SOC \cite{rashba_orbital_2003, flindt_spin-orbit_2006, golovach2006, nowack_coherent_2007, rashba_theory_2008, tokura_coherent_2006, yoneda2018}. Interestingly, the longitudinal effective magnetic field induced by the s-SOC opens a new avenue for spin manipulation, allowing EDDSR. Moreover, it is possible to allow the ``ultra-strong'' EDDSR of a single-electron dressed-spin qubit, where high-harmonic resonances exist \cite{stehlik_extreme_2014, stehlik_role_2016}.

Suppose a two-tone (rather than single tone) oscillating electric field $\vec{E}(t)$ is applied along the $x$-axis, $\vec{E}(t) = \sum_{k=1,2}\vec{E}_{k,max}\cos(\omega_k t + \phi_k)$,
where $k=1,2$ are for a microwave and a radio frequency (rf) components, $\vec{E}_{k,max}$ is the field magnitude, and $\omega_k$ and $\phi_k$ are the frequency and the phase of the field.
Driven by this electric field, the electron spin experiences an effective oscillating magnetic field via the s-SOC.
The driven spin Hamiltonian takes the form
\beq
H_0=\frac{\omega_Z}{2}\sigma_Z + \sum_k (\Omega_{kt}\sigma_X+ \Omega_{kl} \sigma_Z) \cos(\omega_k t+\phi_k),
\eeq
where $\Omega_{kt} = - g\mu_B {b}_{1t}\frac{eE_{k,max}}{2m^*\omega_d^2}$ and $\Omega_{kl} = - g\mu_B {b}_{1l}\frac{eE_{k,max}}{2m^*\omega_d^2}$ are the maximum transverse and longitudinal magnetic field the spin experiences.  While the transverse component $\Omega_{1t}$ is normally used as the EDSR for spin manipulation, below we show that the longitudinal component $\Omega_{2l}$ provides a channel for the manipulation of a dressed-spin qubit.

\begin{figure}[t]
\includegraphics[scale=0.35]{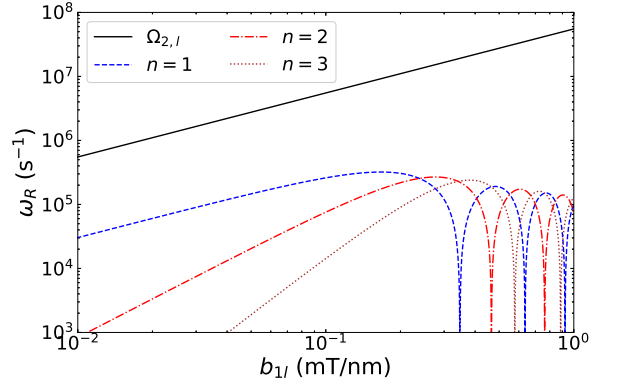}
\caption{
Rabi frequencies of the EDDSR of a dressed-spin qubit versus the longitudinal field gradient $b_{1l}$. The solid line is the Rabi frequency when $\Delta=0$; While the rest are for the $n$-th harmonic resonance when $\Delta=5\times10^6$ $s^{-1}$. 
High harmonic spin resonances are possible in the presence of the longitudinal field gradient $b_{1l}$.
The Rabi frequencies $\Omega_{2l}$ and $\omega_R(n=1)$ increase linearly with $b_{1l}$; While $\omega_R(n=2)$ and $\omega_R(n=3)$ increase quadratically and cubically with $b_{1l}$.
}\label{spinManipLong_bx}
\end{figure}

When the microwave frequency $\omega_1$ is nearly resonant with the spin Larmor frequency $\omega_Z$, i.e. $|\omega_1-\omega_Z|\ll \omega_1$, the dressed qubit in the rotating frame that rotates at $\omega_1$ is governed by
\beq
H_{0}^\prime\approx\frac{\Delta}{2}\sigma_Z + \Omega_{1t}\sigma_{X^\prime}
 + \Omega_{2l} \sigma_Z \cos(\omega_2 t+\phi_2),
\eeq
where $\Delta=\omega_Z-\omega_1$ is the microwave detuning from the Zeeman frequency, axis $X^\prime$ rotates about $Z$ axis at frequency $\omega_1$, and the dressed-spin qubit is defined by the eigenstates of $H_{0,DQ}^\prime = \frac{\Delta}{2}\sigma_Z + \Omega_{1t}\sigma_{X^\prime}$ in the rotating frame. Notice that the $\Omega_{1l}$ term has been omitted since its effect is negligible when $\omega_1 \gg \Omega_{1t},\Delta$.  The $\Omega_{2t}$ term has also been omitted since $\omega_1 \pm \omega_2$ is always far off-resonance from the frequency of the dressed-spin qubit. Clearly, the longitudinal rf field now drives the Rabi oscillation (i.e. EDDSR) of the dressed-spin qubit.
In particular, when $\Delta=0$, the Rabi frequency of the dressed-spin qubit is $\omega_{R}=\Omega_{2l}$, where the resonance condition for the rf field is $\omega_{2}=\Omega_{1t}$ \cite{laucht_breaking_2016, laucht_dressed_2017}. These conditions provide direct experimental access to $\Omega_{2l}$ and $\Omega_{1t}$, and could be a very precise approach for characterizing the magnetic field gradient in the system, making the dressed qubit a sensitive probe of the local magnetic field gradients.

With the EDDSR driving, the system can in principle reach the ``ultra-strong'' driving regime for the dressed qubit, where $\Omega_{2l}\gg \max(\Delta, \Omega_{1t}$, $\omega_{2})$.
Suppose $\omega_{2} > \Omega_{1t}$, after a unitary transformation $H^\prime = U^\dag H U$ with $U = e^{i\frac{\Omega_{2l}}{\omega_{2}} \sigma_Z \sin(\omega_{2}t+\phi_{2})}$, the Hamiltonian becomes 
\beq
H_{0}^{\prime\prime} \approx \frac{\Delta}{2}\sigma_Z + \sum_n \Omega_{R,n}(\sigma_+ e^{-in(\omega_{2}t+\phi_2)} +h.c.), 
\label{eq:driven_dressed}
\eeq
where $\Omega_{R,n}=\Omega_{1t}J_n\left(\frac{\Omega_{2l}}{\omega_{2}}\right)$ and $J_n(x)$ is the Bessel functions of the first kind.  When $\Delta \gg \Omega_{1t}$, the resonance condition becomes $n\omega_{2}=\Delta$, and the Rabi frequency is $\omega_R=\Omega_{R,n}$ for the $n$-th harmonic resonance.

Figure \ref{spinManipLong_bx} shows the Rabi frequency of the EDDSR of a dressed-spin qubit as a function of the longitudinal gradient $b_{1l}\equiv \partial B_x/\partial x$ when $\Delta=0$ (black solid line). The red dashed, blue dotted, and brown dash-dotted lines give the Rabi frequencies for n = 1, 2, and 3 harmonic resonances when $\Delta = 5 \times 10^6$ $s^{-1}$.  When $\Delta = 0$ (resonant microwave driving), the Rabi frequency $\omega_{R}=\Omega_{2l}$ shows the normal linear dependence on the driving field amplitude, which is proportional to the longitudinal gradient $b_{1l}$, with $\Omega_{2l}\propto b_{1l}$.  When $\Delta = 5\times 10^6$ $s^{-1}$ so that $\omega_{2} \approx \Delta/n \gg \Omega_{1t}$ ($\Omega_{1t} \sim 5\times 10^5$ $s^{-1}$ is chosen), higher harmonic resonances become possible, indicating that the system has reached the regime of ultra-strong driving for the dressed-spin qubit.  In this case, for the $n=1$ harmonic, the Rabi frequency initially shows a linear $b_{1l}$ dependence, then oscillates with a period determined by $\Omega_{2l}(b_{1l})\approx \Delta$ as $b_{1l}$ increases. For the $n=2$ ($n=3$) harmonic, the Rabi frequency shows a $b_{1l}^2$ ($b_{1l}^3$) dependence initially, then oscillates as $b_{1l}$ further increases.

Physically, these nonlinear features are a result of the so-called coherent destruction of tunneling, or Landau-Zener-St\"{u}ckelberg (LZS) interference \cite{grossmann1991,grifoni_driven_1998, ashhab2007,shevchenko_landauzenerstuckelberg_2010}.
Among the different multiphoton resonances in Fig.~\ref{spinManipLong_bx}, a maximum Rabi frequency of $\omega_{R}=10^5$ $s^{-1}$ can be reached.  This is faster than the spin dephasing rate, so that such multiphoton resonances are not washed out by dephasing. In addition, the maximum speed of the Rabi oscillation for each multiphoton resonance is achieved at a certain $b_{1l}$, which also defines optimal field gradients for fast manipulation of a dressed-spin qubit when $\Delta\ne 0$.




%
\begin{figure}[t]
\includegraphics[scale=0.35]{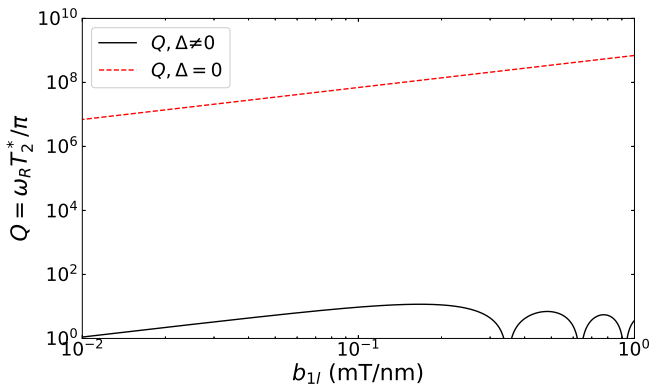}
\caption{
Quality factor $Q=\omega_R T_\varphi/\pi$ of the EDDSR of a dressed-spin qubit versus the magnetic field gradient $b_{1l}$ when $\Delta = 5 \times 10^6$ $s^{-1}$ (black solid) or $\Delta=0$ (red dashed). Because of the suppression of charge noise induced pure dephasing of a dressed qubit, the quality factor of EDDSR can be improved by orders of magnitude when the on-resonance condition $\omega_1=\omega_Z$ is satisfied.
}\label{Qfactor_bx}
\end{figure}

Figure \ref{Qfactor_bx} further shows the quality factor $Q=\omega_RT_2^*/\pi$ for the first harmonic resonance of the EDDSR as a function of the longitudinal gradient $b_{1l}\equiv \partial B_x/\partial x$ when $\Delta = 5 \times 10^6$ $s^{-1}$ (black solid line) and $\Delta=0$ (red dashed line). When $\Delta = 5\times 10^6$ $s^{-1}$, the quality factor of the EDDSR shows nonlinear dependence with the field gradient due to the "ultra-strong" EDDSR driving of a dressed-spin qubit.
When $\Delta = 0$ (i.e. $\omega_1=\omega_Z$), the quality factor $Q$ shows the linear dependence on the longitudinal gradient $b_{1l}$. Because of the fast EDDSR and the suppression of charge noise induced pure dephasing of a dressed qubit, the quality factor of EDDSR is improved by orders of magnitude, which holds promise for applications in quantum information science.

The ultra-strong EDDSR driving could allow study of the effects of ultra-strong driving on a single spin qubit, and examination of LZS interferometry of a single spin \cite{shevchenko_landauzenerstuckelberg_2010}.
Moreover, the realization of the high quality EDDSR provides a new approach for spin manipulation especially when dressed state is used for the selective coupling of qubits \cite{nichol_high_fidelity_2017} or coherence protection \cite{jing2014, miao_universal_2020}.

\section{dressed-spin qubit proposal for quantum computing}
The dressed states together with the EDDSR discussed above can be used as a dressed-spin qubit.
In certain situations, such a dressed qubit can help improve system coherence, while maintaining sizable two-qubit coupling strength \cite{nichol_high_fidelity_2017}.
For instance, the $\sigma_X\sigma_X$ coupling is averaged out in the lab frame when the two qubit frequencies are off-resonance, but remains finite between the dressed qubits. 
Moreover, a dressed qubit can provide coherence protection of spin qubit against $1/f$ charge noise, similar to the coherence protection of dressed states in silicon carbide shown in recent experiment \cite{miao_universal_2020}.
Because of the benefits, we present a proposal for quantum computing based on a dressed-spin qubit in a QD.

Let's consider the initialization a dressed qubit when $\omega_1=\omega_Z$. Suppose the eigenstates of a non-driven spin are $\ket{\uparrow}$ and $\ket{\downarrow}$, then, the two eigenstates of a dressed qubit are $\ket{0}=(\ket{\downarrow} + \ket{\uparrow} e^{i\omega_1 t})/\sqrt{2}$ and $\ket{1}=(\ket{\downarrow} - \ket{\uparrow} e^{i\omega_1 t})/\sqrt{2}$ with $2\Omega_{1t}$ being the energy splitting. 
One way to initialize a dressed qubit is to first initialize the spin to the lower eigenstate $\ket{\downarrow}$ of a non-driven spin as usual \cite{veldhorst2015}. Then, a continuous EDSR driving is applied at Zeeman frequency $\omega_Z$ to dress the spin qubit. 
At the time $t_1=(2n+1/2)\pi/(2\Omega_{1t})$ after the driving, the spin is rotated to state $\ket{0_{Y'}}=(\ket{\downarrow} + i \ket{\uparrow})$. Then, a $\pi/2$ pulse using EDDSR (via the longitudinal gradient) will rotate the state about the $X^{\prime\prime}$-axis, which initializes the spin to the lower eigenstate $\ket{0}$ of a dressed qubit. 

For the readout of a dressed qubit, it is essential to achieve the rotation from the eigenstate $\ket{0}$ of a dressed qubit to the state $\ket{\downarrow}$. Thus, one first rotate the spin about $Z$ by a fast $\pi/2$ EDDSR pulse, which rotates $\ket{0}$ to $\ket{0_{Y'}}$, then, a $\pi/2$ free evolution of the dressed qubit will rotate about the $X^\prime$ axis from $\ket{0_{Y'}}$ to $\ket{\downarrow}$. Then, a measurement of the state $\ket{\downarrow}$ would corresponds to a measurement of the state $\ket{0}$ of the dressed qubit.

The two-qubit gate operation can be achieved by using the exchange interaction $H_{2Q}= J\vec\sigma_{Q1}\cdot \vec\sigma_{Q2}$. Suppose the Zeeman frequency of each spin is  $\omega_{Z,Q1}$ or $\omega_{Z,Q2}$, and microwave of frequencies $\omega_{1,Q1}$ and $\omega_{1,Q2}$ ($\phi_{1,Q1}=\phi_{1,Q2}$ assumed for simplicity) drives the spin qubits into dressed states.
Let's consider $\omega_{1,Q1}=\omega_{Z,Q1}$, $\omega_{1,Q2}=\omega_{Z,Q2}$, and $\Omega_{1t,Q1}=\Omega_{1t,Q2}$, where the dephasing due to $1/f$ charge noise is strongly suppressed and the splitting of the two dressed qubits are on resonance. Then, in the dressed state basis, the interaction Hamiltonian is $H_{2Q}= J(\sigma_{Z^{\prime\prime}}\sigma_{Z^{\prime\prime}} + \sigma_{Y^{\prime}}\sigma_{Y^{\prime}} + \sigma_{X^{\prime\prime}}\sigma_{X^{\prime\prime}})$ \cite{nichol_high_fidelity_2017}. 
In the presence of the exchange interaction, a two-qubit $\sqrt{\text{SWAP}}$ gate can be realized \cite{loss1998}.
Once the $1/T_{\varphi,DQ}$ is suppressed for a dressed qubit, such as a hundred time decrease due to the coherence protection, a smaller $J$ that is bigger than $1/T_{\varphi,DQ}$ can be chosen, where the dephasing due to the exchange interaction can be substantially reduced (also a hundred times decrease) \cite{hu2006,huang_spinDephasing_2018}.
Thus, the two-qubit gate fidelity of dressed qubits can be greatly enhanced due to the improved the coherence time of a dressed qubit. 

By combining the two-qubit gate and the EDDSR of the dressed qubits, universal quantum operation can be realized, where both the single- and two-qubit gate fidelity can be greatly improved due to the coherence protection of a dressed qubit.
In the supplementary material, we also investigate the possibility of realizing the strong coupling between a dressed qubit and a superconducting resonator, and the consequent readout and coupling schemes based on the strong coupling to a superconducting resonator.
The coherence protection of a dressed qubit suggests that a dressed qubit together with the EDDSR and exchange coupling may serve as an attractive platform for quantum computing.

\section{Conclusion}

In conclusion, we study the relaxation, pure dephasing, and electrical manipulation of a dressed-spin qubit of a single electron in a quantum dot. We find that the pure dephasing of a dressed qubit due to charge noise can be suppressed substantially at a sweet spot $\omega_1=\omega_Z$. Away from the sweet spot, the dephasing is similar to the case of static spin. Secondly, the relaxation exhibit non-monotonic behavior in contrast to a static spin qubit. In particular, the spin relaxation of a dressed qubit shows $B_0^5$ dependence with the quantizing magnetic field $B_0$ when the Zeeman frequency is big $\omega_Z> \omega_1$, similar to a static spin qubit. However, spin relaxation reaches minimum when $\omega_Z=\omega_0$. At lower $B_0$ field when $\omega_Z<\omega_0$, the spin relaxation is the same as the relaxation at $B_0^\prime=2\omega_0 - \omega_Z$. This is in contrast to a static spin qubit due to the energy compensation from the driving field that drives a static spin to dressed states.
Moreover, we show that the longitudinal component of the spin and electric field, arising from the breaking time-reversal symmetry of the s-SOC, could provide fast EDDSR of a dressed qubit. Based on the slow dephasing and fast EDDSR, we show that the control quality factor of a dressed qubit can have orders of magnitude improvement. Finally, we propose a dressed-spin-based quantum information processing schemes for potential applications in semiconductor quantum computing.


\appendix

\section{Control and coupling scheme for a dressed-spin qubit based on the spin-photon coupling}

%
For the initialization and readout of a dressed qubit, we can also make use of the strong longitudinal coupling between a spin qubit and a superconducting resonator. For example, in the dispersive limit ($|\omega_0-\tilde\Delta|\gg g_{s,t}$, where $\tilde\Delta=\sqrt{\Delta^2+4\Omega_{1t}^2}$), the state of the dressed qubit can be inferred from the frequency shift of the resonator when the strong coupling limit ($g_{s,t}> 1/T_1, 1/T_2, \kappa$) is achieved, where $\kappa$ is the decay rate of the superconducting resonator.
Moreover,
when the transverse driving provides the quantization axis of a dressed qubit, the EDDSR (via longitudinal magnetic field gradient) will provide an extra axis for spin rotation.

One can also couple the spin qubits via the coupling to a superconducting resonator \cite{huang_impact_2020}. In the presence of microwave driving, the effective two-qubit Hamiltonian in the rotating frame takes the form 
$H_{2Q,DQ}=\sum_i \Delta_{i} \sigma_{Z,i}/2 + \sum_i \Omega_{1t,i} \sigma_{X,i}
+ J_{ZZ}\sigma_{Z,i}\sigma_{Z,i}/2 + J_{XX}(\sigma_{+,1}\sigma_{-,2}+\sigma_{-,1}\sigma_{+,2})
+\sum_i \Omega_{2l,i} \sigma_{Z,i}
\cos\omega_2 t$. When the frequency of transverse EDSR field on each qubit is near resonance with the microwave driving field ($|\Delta_{i}| = |\omega_{Z,i}-\omega_1|\ll \Omega_{1t,i}$), the Hamiltonian in the rotating frame is
\beqa
H_{2Q,DQ} &=&\sum_i \Omega_{1t,i} \sigma_{X,i} + \sum_i \Omega_{2l,i} \sigma_{Z,i}\cos\omega_2 t
\nonumber\\
&+& \frac{J_{ZZ}}{2}(\sigma_{d+,1}\sigma_{d-,2} + h.c.) + \frac{J_{XX}}{2}\sigma_{X,1}\sigma_{X,2},
\eeqa
where $\sigma_{d\pm,i}$ are the creation and annihilation operators defined for the dressed qubits.
Therefore, even when the bare spin splittings $\omega_{Z,i}$ are different for the two qubits, if the near resonance condition $|\omega_{Z,i}-\omega_1|\ll \Omega_{1t,i}$ is satisfied, $J_{XX}\sigma_X\sigma_X$ coupling can be realized, which can be used for the CZ gate (spin quantization here is along $X$-axis due to the strong on-resonance transverse driving). Furthermore, in the absence of longitudinal EDSR, iSWAP gate can be realized by $J_{ZZ}$ term. In the presence of the longitudinal EDSR and $J_{XX}$ coupling, a resonant CNOT gate for dressed qubits can be realized via selective spin rotation \cite{zajac_resonantly_2018}. We emphasize that the above analysis is valid for any two qubits among multiple qubits of different qubit splitting that are coupled to a resonator. Therefore, by using a dressed qubit, it is possible to achieve a selective two-qubit gate among multiple qubits, the sizable $J_{XX}$ coupling can be utilized for fast two-qubit gate (without being averaged out to zero if the frequencies of the two qubits are off-resonance), and relaxes the requirement of tuning the splittings of both qubits into resonance with a resonator, which can be a challenging task in experiments \cite{borjans_resonant_2020}.


\bibliographystyle{apsrev4-1}
\bibliography{spinMFG}

\end{document}